\documentclass[12pt]{article}
\usepackage{natbib}
\usepackage[margin=1in]{geometry}
\usepackage{amsmath, amssymb}
\usepackage{graphicx}
\usepackage{authblk}
\usepackage{setspace}
\usepackage{multirow}
\usepackage{booktabs}
\usepackage{endfloat} 
\usepackage{caption}
\usepackage{subcaption}
\begin{document}

\title{Penalized Distributed Lag Interaction Model: Air Pollution, Birth Weight and Neighborhood Vulnerability}

\author[1]{Danielle Demateis}
\author[1]{Kayleigh P. Keller}
\author[2]{David Rojas-Rueda}
\author[3]{Marianthi-Anna Kioumourtzoglou}
\author[1,*]{Ander Wilson}
\affil[1]{Department of Statistics, Colorado State University, Fort Collins, CO, USA}
\affil[2]{Department of Environmental and Radiological Health Sciences, Colorado State University, Fort Collins, CO, USA}
\affil[3]{Department of Environmental Health Sciences, Columbia University Mailman School of Public Health, New York, NY, USA}
\affil[*]{ander.wilson@colostate.edu}
\date{}
\maketitle

\begin{abstract}
Maternal exposure to air pollution during pregnancy has a substantial public health impact. Epidemiological evidence supports an association between maternal exposure to air pollution and low birth weight. A popular method to estimate this association while identifying windows of susceptibility is a distributed lag model (DLM), which regresses an outcome onto exposure history observed at multiple time points. However, the standard DLM framework does not allow for modification of the association between repeated measures of exposure and the outcome. We propose a distributed lag interaction model that allows modification of the exposure-time-response associations across individuals by including an interaction between a continuous modifying variable and the exposure history. Our model framework is an extension of a standard DLM that uses a cross-basis, or bi-dimensional function space, to simultaneously describe both the modification of the exposure-response relationship and the temporal structure of the exposure data. Through simulations, we showed that our model with penalization out-performs a standard DLM when the true exposure-time-response associations vary by a continuous variable. Using a Colorado, USA birth cohort, we estimated the association between birth weight and ambient fine particulate matter air pollution modified by an area-level metric of health and social adversities from Colorado EnviroScreen.
\vspace{1em}

\end{abstract}

\section{Introduction}

A growing body of epidemiological research supports an association between maternal exposure to air pollution and birth and children's health outcomes  \citep{sram_ambient_2005, bosetti_ambient_2010, stieb_ambient_2012, lakshmanan_associations_2015, jacobs_association_2017}. One goal of such studies is to estimate the exposure-time-response function, which quantifies the association between exposures at multiple time points and a subsequent outcome. Another goal is to identify windows of susceptibility---periods during gestation when increased exposure can alter fetal development. Previous studies have estimated the exposure-time-response function and identified critical windows of susceptibility to air pollution associated with increased risk for asthma \citep{hsu_prenatal_2015}, premature birth \citep{warren_spatial-temporal_2012, chang_assessment_2015}, low birth weight \citep{warren_air_2013}, and other adverse outcomes \citep{chiu_prenatal_2016}. Most studies focused on estimating the exposure-time-response relationship and windows of susceptibility assume homogeneity across the population. However, both the toxicity of chemical exposure and the timing of the windows of susceptibility may vary by individual- or neighborhood-level factors \citep{becklake_gender_1999, lakshmanan_associations_2015, wilson_bayesian_2017, lee_prenatal_2020, niu_association_2022}. Some studies have aimed to address this heterogeneity with models stratified by a single categorical factor such as child sex \citep{bose_prenatal_2017, lee_prenatal_2018, bose_prenatal_2018}. However, there are still gaps in statistical methodology aiming to identify windows of susceptibility and estimate the exposure-time-response relationship with modification by continuous individual- and neighborhood-level variables. 

Distributed lag models (DLMs) are a popular method to estimate the association between an exposure observed longitudinally, such as weekly or daily measures of exposure during pregnancy, and birth and children's health outcomes \citep{hsu_prenatal_2015, chiu_prenatal_2016, chiu_prenatal_2017, bose_prenatal_2017}. The objective of a DLM analysis is to simultaneously estimate the exposure-response relationship at each of the exposure time points, i.e., the exposure-time-response relationship. Compared to using exposure averaged over pre-specified windows, such as clinically defined trimesters, a DLM allows for data-driven identification of windows of susceptibility and has reduced bias for the time-specific exposure-response association \citep{wilson_potential_2017}. A standard DLM analysis assumes a linear exposure-response relationship between each exposure time and the outcome \citep{schwartz_distributed_2000, zanobetti_generalized_2000}. The linear effect of exposure at each time point is usually constrained to vary smoothly across exposure times to reduce variance of the estimator that results from high autocorrelation between repeated measures of exposure. 

Many authors have proposed extensions to DLMs adding flexibility to address specific research questions or improve performance. \cite{armstrong_models_2006} and \cite{gasparrini_distributed_2010} proposed the distributed lag non-linear model, which is of particular interest to this paper. Distributed lag non-linear models relax the linear assumption for DLMs by introducing a cross-basis, a bi-dimensional representation of both the exposure time and the exposure values. While this allows for non-linear exposure relationships at different time points, both the standard DLM and distributed lag non-linear model frameworks assume homogeneous associations across the population. 

To estimate modification in the distributed lag model framework, \cite{wilson_bayesian_2017} proposed a Bayesian distributed lag interaction model (BDLIM). BDLIMs allow for interaction between repeated measures of exposure and a single categorical factor within the DLM framework. However, this approach is limited by a small set of categorical variables or by stratifying continuous variables to use as the effect modifier. \cite{warren_spatially_2020} developed a spatially-varying Gaussian process model for windows of susceptibility, which allows the exposure-time-response function to vary across areal units (e.g., census tract). However, this approach does not attribute variation in the exposure-time-response function to any specific area-level modifying variable. \cite{mork_heterogeneous_2023} proposed a heterogeneous DLM that extends the DLM framework using Bayesian additive regression trees to estimate different exposure-time-response functions for various sub-groups across many covariates. The heterogeneous DLM allows for increased flexibility with multiple continuous and categorical modifiers, but it comes at the cost of increased computational expense, is more challenging to interpret the resulting posterior sample, and does not impose a smoothness constraint on the modifying factor.  Existing methodology is clearly lacking a flexible and computationally efficient method to assess modification attributable to a single continuous modifying variable.

Previous studies have aimed to identify effect modification based on a single continuous variable but were forced to choose alternative approaches due to the lack of methods for a single continuous modifier. One approach is discretizing the continuous modifier into categories. For example, \cite{niu_association_2022} used BDLIMs to investigate the association between air pollution and birth weight stratified by individual stress measurements and the California EnviroScreen composite score, a neighborhood-level measure of burden from pollution and population vulnerability. Several other studies have used BDLIMs with dichotomized versions of continuous predictors, including antioxidant intake \citep{lee_prenatal_2020} and maternal stress \citep{bose_prenatal_2017, lee_prenatal_2018}. Such an approach is limited in identifying windows of susceptibility by pre-determined strata of the study population. Another approach is using time-averaged measures of exposure (e.g., pregnancy average exposure) modified by a continuous factor.  \cite{martenies_associations_2022} use a neighborhood-level score modeled after the California EnviroScreen score as a continuous modifier but fit a linear regression model using annual average exposure. The aggregation in this approach prevents identifying windows of susceptibility during pregnancy. Neither of these approaches can identify windows of susceptibility and estimate modification by a continuous variable. 

We investigate the potential for the association between maternal exposure to ambient air pollution and birth weight to be modified by area-level factors that may influence vulnerability. Evidence shows that a myriad of neighborhood-level factors (e.g., gentrification, poverty rate, green space, residential segregation, and healthcare accessibility) can modify the association between environmental exposures and health outcomes \citep{arcaya_research_2016, schnake-mahl_gentrification_2020, martenies_using_2022}. Estimating the modification of many neighborhood factors is challenging because neighborhood-level factors can be highly correlated with each other, making it difficult to differentiate the modification attributable to separate correlated factors. Therefore, several studies have used an index aggregating multiple neighborhood factors as a candidate modifying factor to study the effect of air pollution on health. Most of these studies considered averaged exposure \citep{martenies_associations_2022}, while others have used DLMs for repeated measures of exposure stratified by a dichotomized version of a composite index \citep{niu_association_2022}. We consider the Colorado EnviroScreen health and social factors score, which is a measure of neighborhood-level health and social vulnerabilities in Colorado, as a modifier of the association between ambient air pollution and birth weight \citep{colorado_department_of_public_health_and_environment_cdphe_colorado_2022, colorado_EnviroScreen_tool_team_colorado_2022}.

In this paper, we propose a distributed lag interaction model that allows each individual to have a personalized exposure-time-response function based on a continuous modifying variable. Unlike the standard DLM, our model can estimate unique effects for each possible value of the modifying factor across all exposure time points. The proposed approach builds on the cross-basis approach for the distributed lag non-linear model framework of \cite{gasparrini_distributed_2010} but adapts it to capture the modified exposure-time-response relationship. Specifically, the approach uses basis expansions in two dimensions. Expansion in the modifier dimension allows the exposure-time-response function to vary for different levels of the modifier. Expansion in the exposure-time dimension constrains the exposure-time-response function to vary smoothly across exposure time points, providing the necessary regularization in the presence of multicollinearity between repeated measures of exposure. We estimated the association between maternal exposure to ambient fine particulate matter ($\text{PM}_{2.5}$) and birth weight using a large administrative data set derived from the Colorado Birth Registry. In our analysis, we accounted for modification with the Colorado EnviroScreen health and social factors score to assess how maternal neighborhoods influence vulnerability to ambient $\text{PM}_{2.5}$. We identified windows of susceptibility and were also able to identify neighborhoods with heightened sensitivity. Our software is available in the \texttt{R} package \texttt{dlim}. The package is publicly available on GitHub.

\section{Colorado Birth Cohort Data} \label{sec.data}

\begin{figure}
    \centering
    \begin{subfigure}[b]{0.49\textwidth}
        \centering
        \includegraphics[width=1\textwidth]{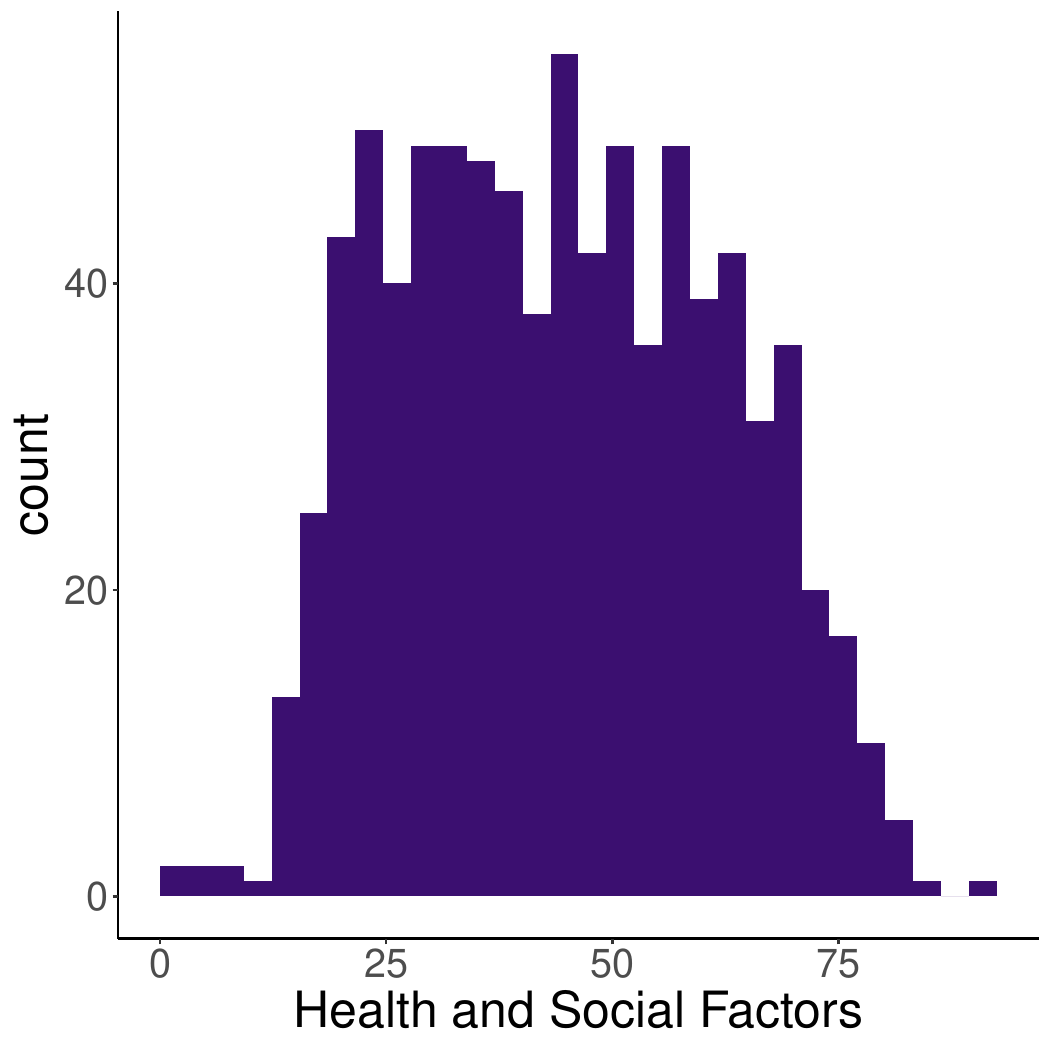}
    \end{subfigure}
    \hfill
    \begin{subfigure}[b]{0.49\textwidth}
        \centering
        \includegraphics[width=1\textwidth]{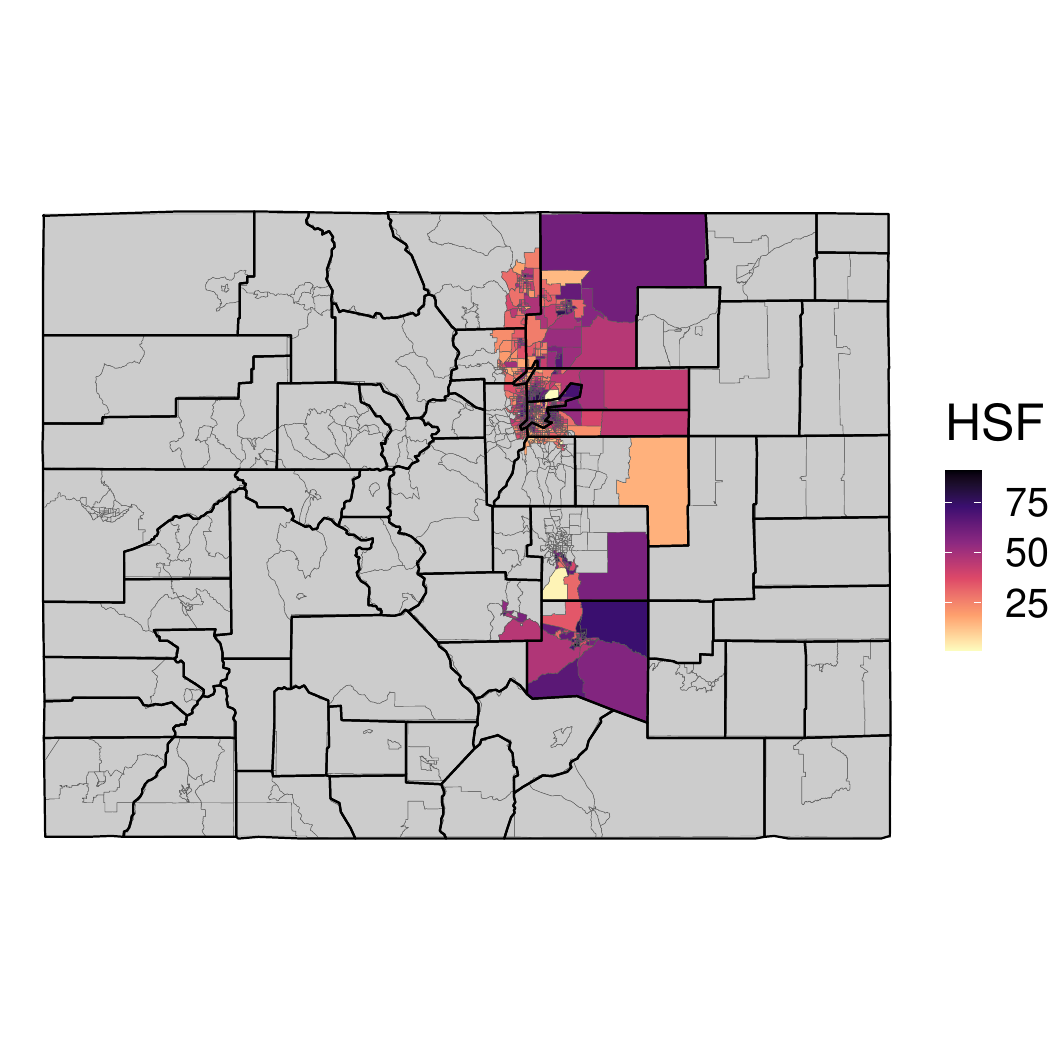}
    \end{subfigure}
    \caption{Summaries of Colorado EnviroScreen health and social factors (HSF) scores for census tracts selected in this study. Histogram of HSF scores for census tracts included in this study, with quartiles at scores 32, 46, and 59 (left). Map of the state of Colorado shaded by census tract. Tracts not included in our study are shaded in gray (right).}
    \label{fig:HSF}
\end{figure}

Our Colorado birth cohort data set contains all live births in Colorado, USA with dates of birth between 2007 and 2018. We limited our analysis to full-term (estimated gestational age of 37 weeks or longer), singleton births with estimated dates of conception between 2007 and 2017. Exposure measurement error may be greater in the mountainous regions of the state, and confounding by elevation is associated with lower birth weight \citep{yip_altitude_1987}. To reduce the impact of both exposure measurement error and confounding by elevation, we further limited to births in counties constituting Colorado's Front Range region (extending east of the Rocky Mountains from Colorado Springs to the Wyoming boarder) and in census tracts at elevation, based on the centroid, of 6000 feet or lower (Figure~\ref{fig:HSF}).   

The data set includes birth weight, along with measurements before pregnancy of the pregnant person’s age, body mass index (BMI), height, weight, race and ethnicity, education, income, and marital status. The data set also includes maternal self-reported prenatal care habits and smoking habits during pregnancy. See Supplemental Table 9 for more information on these variables. The outcome of interest is birth weight for gestational age z-score (BWGAZ) constructed using the Fenton growth charts \citep{fenton_new_2003}.

We obtained environmental justice scores from Colorado EnviroScreeen on the census-tract level \citep{colorado_EnviroScreen_tool_team_colorado_2022, colorado_department_of_public_health_and_environment_cdphe_colorado_2022}. Colorado EnviroScreen synthesizes environmental exposures and effects, climate vulnerability, sensitive populations, and demographics across multiple years to create a proxy for long-term environmental injustice in an area. Scores are percentiles, ranging from 0 to 100, with a higher score indicating that the area is more likely to experience environmental health injustices. We used the health and social factors score, which quantifies a population's long-term susceptibility to environmental changes and vulnerability to social or economic adversities in an area. The health and social factors score comprises two component scores: a sensitive population score and demographics score. The sensitive population component score is an aggregation of variables on the census-tract level: asthma hospitalization rate, prevalence of cancer and diabetes, rate of heart disease in adults, life expectancy, rate of low birth weight, prevalence of poor mental health conditions, and the percent of elderly and young children for an area. The demographics component score combines the percent of the total population in a census tract that are burdened by housing costs, have disabilities, have less than high school education, have limited English proficiency, are people of color, and have low income. We linked the EnviroScreen health and social factors score to the Colorado birth cohort based on census tract of maternal residence at birth.

We also obtained daily ambient $\text{PM}_{2.5}$ concentrations for each census tract from the down-scaled models published by the US Environmental Protection Agency, which combine measurements from regulatory monitoring networks and output from atmospheric chemistry models (https://www.epa.gov/hesc/rsig-related-downloadable-data-files). Using the census tract of maternal residence at birth, we linked exposure data to each birth and computed average exposure for each week of gestation. We also linked ambient daily maximum temperature at the census-tract level from \cite{abatzoglou_development_2013}. Our final analysis data set included 393,205 births in 786 census tracts across 13 counties.

\section{Modeling Framework} \label{sec. framework}

\subsection{Model Approach} \label{sec. model}
For individual $i = 1 \dots n$, $y_i$ is the response, $x_{it}$ is the exposure at time $t$ for $t = 1, \dots, T$, and $\mathbf{z}_{i}$ is a covariate vector including the intercept. Assuming $y_i$ follows an exponential family distribution, a standard DLM with no modification is
\begin{equation}\label{eq.DLM}
    g[E(y_i|\mathbf{x}_{i},\mathbf{z}_i)]  = \sum_{t=1}^{T} x_{it}\beta_t +  \mathbf{z}_i'\boldsymbol\gamma,
\end{equation}
where $g$ is the link function, $\beta_{t}$ is the linear effect of exposure at time $t$ on the response, and $\boldsymbol\gamma$ is a vector of regression coefficients. The vector $\boldsymbol\beta=(\beta_1,\dots,\beta_T)’$ is the exposure-time-response function. The exposure-time-response function of a standard DLM is the same for all individuals. 

We are interested in estimating how the exposure-time-response function varies as a function of a scalar, continuous modifying variable $m_i$. Throughout, we assume that $m_i$ is also included in the covariate vector $\mathbf{z}_i$. Our proposed distributed lag interaction model is
\begin{equation}\label{eq.main}
    g[E(y_i|\mathbf{x}_{i},\mathbf{z}_i, m_i)] = \sum_{t=1}^{T} x_{it}\beta_t(m_i) +  \mathbf{z}_i'\boldsymbol\gamma.
\end{equation}
Unlike the standard DLM formulation, the linear effect of exposure on the response at exposure time $t$ for individual $i$, denoted $\beta_t(m_i)$ in model \eqref{eq.main}, depends on the value of the modifier $m_i$ for that individual. In our analysis, $m_i$ is the health and social factors score for the census tract of maternal residence for birth $i$.

We target two parameters from model \eqref{eq.main} for inference. First, we are interested in the exposure-time-response function for the $i^{\text{th}}$ individual with modifier value $m_i$, which is defined as $\boldsymbol\beta(m_i) = [\beta_1(m_i), \dots,  \beta_T(m_i)]'$. The $t^{\text{th}}$ element of $\boldsymbol\beta(m_i)$, $\beta_t(m_i)$, is the linear effect of exposure at time point $t$ for individual $i$ with modifier value $m_i$. Second, we are interested in the cumulative effect of exposure for each individual. In model \eqref{eq.main}, the cumulative effect for individual $i$ is the expected difference in the response associated with one unit higher exposure at all exposure time points, defined as $\delta(m_i) = \sum_{t=1}^T\beta_t(m_i)$.

\subsection{Parameterization of exposure-time-response functions} \label{sec.param}

We allow $\beta_t(m_i)$ in \eqref{eq.main} to vary smoothly in both the exposure-time and modifier dimensions through a penalized spline approach. Following the penalized cross-basis formulation proposed by \cite{gasparrini_penalized_2017}, we create a cross-basis for the modifier-time space. 

To allow for the linear effect of exposure at each time point to vary smoothly as a function of the modifier, we use B-spline basis expansions of the modifier values. The basis expansion for modifier value $m_i$ is denoted $\mathbf{b}(m_i) = [b_1(m_i),\dots,b_{\nu_{mod}}(m_i)]’$, where $\nu_{mod}$ is the user-defined number of basis functions for the modifier basis expansion. We can write $\beta_t(m_i)$ as a linear combination of modifier basis values,
\begin{equation}\label{eq.beta}
    \beta_t(m_i)=\sum_{k=1}^{\nu_{mod}} b_k(m_i)\eta_{tk}.
\end{equation}

Then, we use B-splines to create a basis expansion of the exposure time points to add a smoothness constraint in the time dimension and to regularize the model, reducing the effect of multicollinearity between repeated measures of exposure. The exposure-time basis expansion for time point $t$ is denoted $\mathbf{c}(t) = [c_1(t),\dots,c_{\nu_{time}}(t)]’$, where $\nu_{time}$ is the user defined number of basis functions for the exposure-time basis expansion. We can write $\eta_{tk}$ from \eqref{eq.beta} as a linear combination of exposure-time basis values,
\begin{equation}\label{eq.eta}
    \eta_{tk}= \sum_{j=1}^{\nu_{time}} c_j(t)\theta_{jk},
\end{equation}
where $\theta_{jk}$ for $k = 1, \dots, \nu_{mod}$ and $j = 1, \dots, \nu_{time}$ are regression parameters.

Using the parameterizations \eqref{eq.beta} and \eqref{eq.eta}, the first term on the right-hand side of \eqref{eq.main} can be written in terms of $\{\theta_{jk}\}_{j=1,k=1}^{\nu_{mod},\nu_{time}}$ as
\begin{equation} \label{eq.key}
\begin{split}
 \sum_{t=1}^{T} x_{it}\beta_t(m_i) 
 & = \sum_{k=1}^{\nu_{mod}} \sum_{t=1}^T x_{it}  b_k(m_i) \sum_{j=1}^{\nu_{time}} c_{j}(t) \theta_{jk} \\
 & =  \sum_{j=1}^{\nu_{time}} \sum_{k=1}^{\nu_{mod}} \left[ \sum_{t=1}^T x_{it}  b_k(m_i)  c_{j}(t) \right] \theta_{jk} \\
 & = \sum_{j=1}^{\nu_{time}} \sum_{k=1}^{\nu_{mod}}  w_{jk}(m_i,\mathbf{x}_{i}) \theta_{jk}, 
\end{split}
\end{equation}
where $w_{jk}(m_i,\mathbf{x}_{i})$ is an element of the cross-basis for individual $i$ representing the $j^{\text{th}}$ element in the expansion of the exposure-time basis and the $k^{\text{th}}$ element in the expansion of the modifier basis. Incorporating the representation in \eqref{eq.key} into \eqref{eq.main}, we have the following reparameterized model
\begin{equation}
     g[E(y_i|\mathbf{x}_i,\mathbf{z}_i, m_i)] = \sum_{j=1}^{\nu_{time}} \sum_{k=1}^{\nu_{mod}} w_{jk}(m_i,\mathbf{x}_{i}) \theta_{jk} + \mathbf{z}_i'\boldsymbol\gamma.
\end{equation}
Since $\beta_t(m_i)$ for each $t=1,\dots,T$ is a linear combination of $\{\theta_{jk}\}_{j=1,k=1}^{\nu_{mod},\nu_{time}}$, we obtain all estimates of the exposure-time-response function and cumulative effects using linear transformations of the estimates of $\{\theta_{jk}\}_{j=1,k=1}^{\nu_{mod},\nu_{time}}$ (see Section~\ref{sec.inference} for more details).

\subsection{Parameter Estimation}\label{sec.estimation}

Various basis expansions and estimation procedures can be applied to estimate the model presented in Section~\ref{sec.param}. In this paper, we focus on a penalized spline approach to allow for data-adaptive smoothness in the exposure-time-response surface. We discuss alternative penalty methods in Section \ref{sec: penalty}. Let $l(\boldsymbol\theta,\boldsymbol\gamma)$ be the likelihood for our unpenalized model, where $\boldsymbol\theta = [\theta_{11}, \dots, \theta_{\nu_{time}1}, \theta_{12}, \dots, \theta_{\nu_{time}2}, \dots, \theta_{1\nu_{mod}}, \dots, \theta_{\nu_{time}\nu_{mod}}]'$. Let $\mathbf{D}_{\nu}$ be a $(v-2) \times v$ second order difference matrix and $\mathbf{I}_{\nu}$ be a $\nu \times \nu$ diagonal matrix. Define a $\nu \times \nu$ squared difference matrix $\mathbf{S}^*_{\nu} =\mathbf{D}_{\nu}'\mathbf{D}_{\nu}$. To obtain penalty matrices $\mathbf{S}_{\nu_{time}}$ and $\mathbf{S}_{\nu_{mod}}$, we divide $\mathbf{S}^*_{\nu_{time}}$ and $\mathbf{S}^*_{\nu_{mod}}$ by their first eigenvalues, respectively, to ensure each applies equally weighted penalties. Let $\otimes$ denote the Kronecker product. The penalized likelihood is
\begin{equation}
    l_p(\boldsymbol\theta,\boldsymbol\gamma,\boldsymbol\lambda) = l(\boldsymbol\theta,\boldsymbol\gamma) - \frac{1}{2}\boldsymbol\theta'\{\lambda_{mod} (\mathbf{S}_{\nu_{mod}} \otimes \mathbf{I}_{\nu_{time}}) + \lambda_{time} (  \mathbf{I}_{\nu_{mod}} \otimes \mathbf{S}_{\nu_{time}})\}\boldsymbol\theta,
\end{equation}
where $\boldsymbol\lambda = [\lambda_{mod}, \lambda_{time}]$ are penalty parameters that control smoothness. We fit the model with REML \citep{laird_random-effects_1982} using the \texttt{gam} function from the \texttt{mgcv} package in \texttt{R} \citep{wood_generalized_2017}.

\subsection{Inference} \label{sec.inference}
We conduct inference on both the exposure-time-response function and the cumulative effect for each individual (see Section~\ref{sec. model}). Our estimator of the exposure-time-response function for individual $i$ is $\hat{\boldsymbol\beta}(m_i) = [\mathbf{b}'(m_i) \otimes \mathbf{C} ] \hat{\boldsymbol\theta}$, where $\mathbf{C} = [\mathbf{c}(1) \dots \mathbf{c}(T)]'$. The $t^{\text{th}}$ element of $\hat{\boldsymbol\beta}(m_i)$ can also be written as is $\hat{\beta}_t(m_i) = \sum_{j=1}^{\nu_{time}} \sum_{k=1}^{\nu_{mod}} b_k(m_i) c_j(t) \hat{\theta}_{jk}$. The estimator $\hat{\beta}_t(m_i)$ has variance $V[\hat{\beta}_t(m_i)] = \mathbf{1}_{\nu_{time} \nu_{mod}}' \mathbf{A}_t(m_i) \mathbf{1}_{\nu_{time} \nu_{mod}}$, where 
$\mathbf{A}_t(m_i)$ is the $\nu_{time} \nu_{mod} \times \nu_{time} \nu_{mod} $ matrix
\begin{align}
    \mathbf{A}_t(m_i) =& [\mathbf{c}'(t) \otimes \mathbf{1}_{\nu_{mod} \times \nu_{time} \nu_{mod}}] \odot [\mathbf{1}_{\nu_{time} \times \nu_{time} \nu_{mod}} \otimes \mathbf{b}(m_i)] \\ 
    &\odot [\mathbf{c}'(t) \otimes \mathbf{1}_{\nu_{mod} \times \nu_{time} \nu_{mod}})]' \odot [\mathbf{1}_{\nu_{time} \times \nu_{time} \nu_{mod}} \otimes \mathbf{b}(m_i)]' \odot V[\hat{\boldsymbol\theta}]. \notag
\end{align}
The operator $\odot$ represents element-wise multiplication and $V[\hat{\boldsymbol\theta}]$ is the covariance matrix for the estimated coefficients. We estimate $V[\hat{\boldsymbol\theta}]$ using empirical Bayes estimators within the \texttt{mgcv} package in \texttt{R} \citep{marra_coverage_2012}.

Our estimator of the cumulative effect for individual $i$ is $\hat{\delta}(m_i) = \sum_{t=1}^T\hat{\beta}_t(m_i)$, or equivalently, $\hat{\delta}(m_i) = \mathbf{w}_*'(m_i)\hat{\boldsymbol\theta}$, where $\mathbf{w}_*(m_i)$ is the vectorization by column of $[\mathbf{1}'_{T} \otimes \mathbf{b}(m_i)] \mathbf{C}$. The variance of $\hat{\delta}(m_i)$ is $V[\hat{\delta}(m_i)] = \mathbf{w}_*'(m_i) V[\hat{\boldsymbol\theta}] \mathbf{w}_*(m_i)$.

\subsection{Alternative Approaches for Penalization}\label{sec: penalty}

There are potential alternative penalization structures, several of which are explored in \cite{gasparrini_penalized_2017}. As described in Section~\ref{sec.estimation}, we can impose a penalty in both the modifier and exposure-time dimensions by using difference matrices. The order of the difference matrix is related to the amount of smoothness imposed. In our main approach, we consider second order difference matrix penalization in both dimensions. We could instead use a first order difference matrix in either dimension to impose less smoothness for that particular dimension. We refer to this method as PS($d_{time}, d_{mod}$) penalization, where $d_{time}$ and $d_{mod}$ are the difference matrix orders for the exposure-time and modifier dimensions, respectively.

Another option is to impose penalties on the second derivative in each dimension of the exposure-time-response basis functions \citep{green_nonparametric_1994}. Here, we consider using natural cubic splines with the added constraint of allowing only one spline to be non-zero at each knot for computational convenience \citep{wood_generalized_2017}. We refer to this method as CR penalization.

\subsection{Alternative Approach without Penalization}\label{sec: non-pen}

We can also perform model estimation without penalization using natural splines. However, the use of natural splines requires a priori specification of the degrees of freedom or a data-driven selection process, such as cross-validation, that can be computationally burdensome and difficult in practice. We recommend using AIC to choose degrees of freedom for non-penalized models. 

\subsection{Alternative Approach with Linear Interaction}\label{sec: DLIM-lin}

The proposed model allows for flexible modification of the exposure-time-response function. However, there may be situations where a simpler form of modification is preferable. We can construct the modifier basis using polynomial functions, particularly a polynomial function of degree 1 (i.e., linear), instead of penalized B-splines. The modifier basis expansion for modifier value $m_i$ is then $[1, m_i]$. Using this modifier basis expansion for the cross-basis described in Section~\ref{sec.param} is a distributed lag model with an additional linear interaction term, equivalent to the model $g[E(y_i|x_{it}, m_i, \mathbf{z}_i')] = \sum_{t=1}^{T} x_{it}\beta_t +  \sum_{t=1}^{T} x_{it}m_i\beta_{1t} + \mathbf{z}_i'\boldsymbol\gamma$. We refer to this reduced model as a distributed lag model with linear interaction. Increasing the degree of the polynomial basis for the modifier increases the additional number of interaction terms. Estimation and inference in Sections~\ref{sec.estimation}~and~\ref{sec.inference} proceed the same. We impose two penalty matrices in the lag dimension. The first penalty matrix is $diag(1,0) \otimes \mathbf{S}_{\nu_{time}}$, which penalizes regression coefficient estimates contributing to estimates of the $\beta_t$. The second penalty matrix is $diag(0,1) \otimes \mathbf{S}_{\nu_{time}}$, which penalizes regression coefficient estimates contributing to estimates of the $\beta_{1t}$.

\subsection{Test for Interaction} \label{sec: compare}

To test for effect modification, we developed a bootstrap likelihood ratio test using the approach in \cite{roca-pardinas_testing_2005}. The null model is a standard DLM, as in \eqref{eq.DLM}, a model without effect modification. The full model is a model with effect modification, either our proposed model in \eqref{eq.main} or alternative model presented in Section~\ref{sec: DLIM-lin}. First, we fit the null model and obtain fitted values $\Tilde{y}_i$. For $b=1, \dots, B$, we first generate a sample $y_i^{*b}$ for $i=1,\dots,n$, with $y_i^{*b}$ drawn from the assumed exponential family distribution built around $\Tilde{y}_i$. For a Gaussian model, we sample the residuals ($y_i-\tilde{y}_i$) with replacement and add them to $\Tilde{y}_i$. For non-Gaussian GLM, we draw from the distribution parameterized by $\Tilde{y}_i$ (i.e., $y_i^{*b} \sim \text{Pois}(\Tilde{y}_i)$). We then fit both the full and reduced model to each bootstrap sample and calculate a likelihood ratio test statistic denoted as $\hat{T}^{*b}$. This results in an empirical null distribution for our test. The critical value for specified $\alpha$ level is the $1-\alpha$ quantile of the bootstrap sample of test statistics $\hat{T}^{*1},\dots, \hat{T}^{*B}$. We reject the null hypothesis of no modification when the test statistic fit on the original data is greater than the empirical critical value. We can also reject the null hypothesis based on an empirical bootstrap p-value, which we define as the proportion of bootstrap test statistics that are greater than the test statistic calculated from the original data. We can also apply a similar test to compare the DLIM with nonlinear modification presented in Section~\ref{sec. model} to the distributed lag model with linear interaction presented in Section~\ref{sec: DLIM-lin}.

\subsection{Repeated Measures Among Individuals}\label{sec: random intercept}
We also propose a mixed model representation of the DLIM. For example, one may consider exposure in the weeks prior to each visit in a longitudinal cohort \citep{peng_short-term_2017}. Consider the longitudinal study with individuals $i=1,\dots,N$ and visits $l=1,\dots,n_i$ for individual $i$. We account for repeated measures among individuals by adding an individual-specific intercept $\alpha_i$ to \eqref{eq.main}. The mixed model DLIM is

\begin{equation}
    g[E(y_{il}|\mathbf{x}_{il},\mathbf{z}_{il}, m_{il})] = \sum_{t=1}^{T} x_{il,t}\beta_t(m_{il}) +  \mathbf{z}_{il}'\boldsymbol\gamma + \alpha_i.
\end{equation}

We assume the individual-specific intercepts are independent of each other and the residuals, and are distributed normal with mean zero and constant variance. The modifier $m_{il}$ can be constant over time or time dependent. We fit the model with REML and the \texttt{mgcv} package as described in Section~\ref{sec.estimation}.

\section{Simulations} \label{sec. simulation}

We validated our proposed DLIM with a simulation under four scenarios and three different signal-to-noise ratios. We evaluated the method's ability to estimate the exposure-time-response functions and cumulative effects. We compared the operating characteristics of the proposed model to a DLM  with linear interaction (DLIM-linear) as described in Section~\ref{sec: DLIM-lin} and a standard DLM model with no effect modification. We additionally evaluated the empirical performance of the bootstrap test for modification. We provide results for a DLIM(20,20) with PS(2,2) penalization, i.e., a DLIM with 20 basis functions and penalty with a second order difference matrix in the exposure-time basis and 20 basis functions and penalty with a second order difference matrix in the modifier basis, and a DLIM-linear and standard DLM with nominal 10 degrees of freedom for the exposure-time basis and penalty with a second order difference matrix.  We discuss selection of degrees of freedom and sensitivity in Section~\ref{sec.tuning}, and we show results for models with different number of basis functions and additional models and data generating mechanisms in the supplemental material. Code for implementing these simulations is provided as a companion file to this manuscript.

\subsection{Simulation Design and Data Generation} \label{sec.simDesign}

We randomly generated 1000 birth dates and constructed 37 weeks of exposure data for each birth from observed ambient $\text{PM}_{2.5}$ data. For each birth, we simulated modifier value $m_i$ from a continuous uniform distribution on $(0,1)$ because the modifier of interest in this paper and other motivating papers is a bounded index \citep{martenies_associations_2022, niu_association_2022}. 

\begin{figure}
    \centering
    \includegraphics[width=1\textwidth]{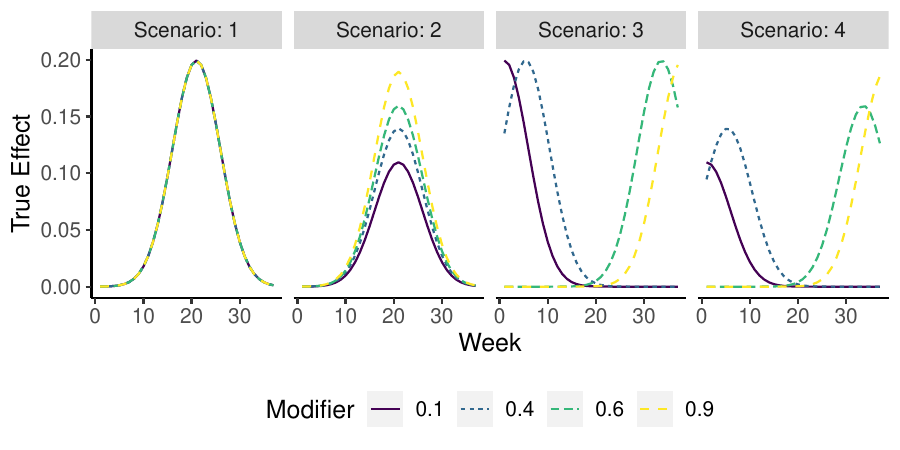}
    \caption{Each curve is an exposure-time-response function for an individual, each with different modifying values. In this example, there are $n=4$ individuals, with modifier values 0.1, 0.5, 0.6, and 0.9, respectively. Each scenario is described in Section~\ref{sec.simDesign}.}
    \label{fig:scenarios}
\end{figure}

We considered four scenarios with increasing complexity of the exposure-time-response function and modification. Figure~\ref{fig:scenarios} illustrates the four scenarios. We based the data generation scenarios on the function $f(t,c) := 2.5\phi[(t-c)/5]$, where $\phi(\cdot)$ is the normal probability density function. We generated data for the scenarios as follows.

\begin{enumerate}
  \item \textbf{No Modification}: The peak effect and its timing are the same for each individual. To simulate this scenario, we let $\beta_t(m_i)=f(t,20)$. 
  \item \textbf{Linear Modification}: The timing of peak effect is the same for each individual, but the peak effect is linearly scaled by each individual's modifying value. To simulate this scenario, we let $\beta_t(m_i)=m_i * f(t,20)$.
  \item \textbf{Non-linear Shift Modification}: The peak effect is the same for each individual but the timing of peak effect is shifted according to a non-linear function of each individual's modifying value. To simulate this scenario, we allowed the centers of the exposure-time-response function normal curves, ranging from 10 to 30, to be a non-linear function of the modifier value, $\beta_t(m_i)=f(t,37[1+\text{exp}\{-20(m_i-0.5)\}]^{-1})$. 
  \item \textbf{Complex Modification}: The peak effect and its timing are functions of each individual's modifying value. To simulate this scenario, we combined scenarios 2 and 3 to obtain $\beta_t(m_i)=m_i*f(t,37[1+\text{exp}\{-20(m_i-0.5)\}]^{-1})$. 
\end{enumerate}

For all scenarios, we generated 3 covariates and their corresponding regression coefficients from a standard normal distribution. To simulate response values, we assumed a Gaussian distribution with variance $\sigma^2$ and used an identity link function $g$ in \eqref{eq.main}. We fixed the regression coefficient for the main effect of the modifier at 1. We investigated model fit for increasing signal-to-noise ratios, including 0.5 (low), 1 (medium), and 10 (high). Here, the signal is the standard deviation of the first term across $i = 1, \dots, n$ in \eqref{eq.main}, and noise is $\sigma$. We simulated 200 data sets for each scenario and signal-to-noise combination.

\subsection{Performance Evaluation}
\label{sec.simEval}

For each simulated data set, we calculated the root mean squared error (RMSE) for the cumulative effect as  $[n^{-1}\sum_{i=1}^n \{\delta(m_i) - \hat{\delta}(m_i)\}^2]^{1/2}$ and cumulative coverage as  $n^{-1}\sum_{i=1}^n \mathbb{I}[ |\delta(m_i) - \hat{\delta}(m_i)| \allowbreak \le \Phi^{-1}(1-\alpha/2) \hat{V}^{1/2}\{ \hat{\delta}(m_i) \}]$. Here, $\mathbb{I}(A) = 1$ if event $A$ occurs and $\mathbb{I}(A) = 0$ if event $A$ does not occur. We also calculated RMSE for point-wise effects as  $[n^{-1} T^{-1} \sum_{i=1}^n    \sum_{t=1}^T \{\hat\beta_t(m_i) - \beta_t(m_i)\}^2]^{1/2}$ and point-wise coverage as $n^{-1} T^{-1} \sum_{i=1}^n  \sum_{t=1}^T \mathbb{I}[ |\beta_t(m_i) - \hat{\beta}_t(m_i)| \le \Phi^{-1}(1-\alpha/2) \hat{V}^{1/2} \{ \hat{\beta}_t(m_i) \}]$. We averaged RMSE and coverage across the 200 simulated data sets. 

To evaluate our testing procedure we conducted three tests on each data set. These tests compared: 1) DLIM(20,20) to DLM, 2) DLIM-linear to DLM, and 3) DLIM(20,20) to DLIM-linear. We report the proportion of times that the null was rejected for each test.

\subsection{Simulation Results}

\subsubsection{Parameter Estimation}\label{sec.parameter estimation results}

Table~\ref{tab: main12} includes results for simulation scenarios 1 and 2, and Table~\ref{tab: main34} includes results for simulation scenarios 3 and 4. Overall, our simulation demonstrates that the DLIM(20,20) can capture complex effect modification patterns. Further, using the DLIM over the standard DLM framework results in substantially less bias and more accurate inference when there is effect modification and only a small loss in efficiency when there is no modification.

\begin{table}[h]
\centering
    \caption{Cumulative and point-wise performance metrics for penalized models averaged over all 200 simulated data sets for scenarios 1 and 2 and three signal-to-noise (SNR) levels. The low SNR is 0.5, medium SNR is 1, and the high SNR is 10. The number of basis functions for the exposure-time basis was 10 for both the DLM and DLIM-linear. The table also includes the proportion of rejection ($\alpha = 0.05$) across simulated data sets under each null model (DLM or DLIM-linear) using the model comparison procedure described in Section \ref{sec: compare}.}
    \begin{tabular} {c c c c c c c c}
        \toprule
        & & \multicolumn{2}{c}{Cumulative} & \multicolumn{2}{c}{Point-wise} & \multicolumn{2}{c}{Model Comparison}   \\ 
        \cmidrule(lr){3-4} \cmidrule(lr){5-6} \cmidrule(lr){7-8}
        SNR & Model &  RMSE & Coverage & RMSE & Coverage & DLM & DLIM-linear  \\ \hline  

        \multicolumn{8}{c}{Scenario 1: No Modification} \\ \hline

        & DLIM(20,20) & 0.263 & 0.95 & 0.026 & 0.92 & 0.04 & 0.05 \\ 
         low & DLIM-linear & 0.257 & 0.95 & 0.024 & 0.92 & 0.06 & -\\ 
         & DLM & 0.169 & 0.96 & 0.020 & 0.91 & - & - \\
         \hline

         & DLIM(20,20) & 0.131 & 0.95 & 0.015 & 0.96 & 0.04 & 0.05\\
         medium & DLIM-linear & 0.129 & 0.95 & 0.013 & 0.94 & 0.05& -\\ 
         & DLM & 0.084 & 0.96 & 0.011 & 0.94& -& - \\  

         \hline
         
         & DLIM(20,20) & 0.013 & 0.95 & 0.003 & 0.98 & 0.04 & 0.04 \\ 
         high & DLIM-linear & 0.013 & 0.95 & 0.002 & 0.90 & 0.04& -\\ 
         & DLM & 0.008 & 0.96 & 0.002 & 0.90 & -& -\\ 

         \hline
         
         \multicolumn{8}{c}{Scenario 2: Linear Modification} \\ \hline
         & DLIM(20,20) & 0.331 & 0.95 & 0.030 & 0.90 & 0.20 & 0.04 \\ 
         low & DLIM-linear & 0.329 & 0.95 & 0.029 & 0.89 & 0.30 & -\\ 
         & DLM & 0.438 & 0.70 & 0.028 & 0.83 & -& -\\ 
         \hline

         & DLIM(20,20) & 0.166 & 0.95 & 0.017 & 0.93 & 0.67 & 0.06\\ 
         medium & DLIM-linear & 0.166 & 0.95 & 0.017 & 0.91 & 0.80& -\\ 
         & DLM & 0.383 & 0.40 & 0.019 & 0.78& -& - \\ 
         \hline
         
         & DLIM(20,20) & 0.017 & 0.95 & 0.003 & 0.98 & 1.00 & 0.05 \\
         high & DLIM-linear & 0.016 & 0.95 & 0.002 & 0.94 & 1.00& -\\ 
         & DLM & 0.361 & 0.06 & 0.014 & 0.55 & -& -\\ 
         \hline
         \bottomrule
         
    \end{tabular} 
    
    \label{tab: main12}   
\end{table}

\begin{table}[h]
\centering
    \caption{Cumulative and point-wise performance metrics for penalized models averaged over all 200 simulated data sets for scenarios 3 and 4 and three signal-to-noise (SNR) levels. The low SNR is 0.5, medium SNR is 1, and the high SNR is 10. The number of basis functions for the exposure-time basis was 10 for both the DLM and DLIM-linear. The table also includes the proportion of rejection ($\alpha = 0.05$) across simulated data sets under each null model (DLM or DLIM-linear) using the model comparison procedure described in Section \ref{sec: compare}.}
    \begin{tabular} {c c c c c c c c}
        \toprule
        & & \multicolumn{2}{c}{Cumulative} & \multicolumn{2}{c}{Point-wise} & \multicolumn{2}{c}{Model Comparison}   \\ 
        \cmidrule(lr){3-4} \cmidrule(lr){5-6} \cmidrule(lr){7-8}
        SNR & Model &  RMSE & Coverage & RMSE & Coverage & DLM & DLIM-linear \\ \hline

         \multicolumn{8}{c}{Scenario 3: Non-linear Shift Modification} \\ \hline
         & DLIM(20,20) & 0.404 & 0.94 & 0.049 & 0.82 & 1.00 & 1.00 \\ 
         low & DLIM-linear & 0.620 & 0.79 & 0.053 & 0.73 & 0.52 & -\\ 
         & DLM & 0.555 & 0.72 & 0.065 & 0.49& -& - \\ 
         \hline

        & DLIM(20,20) & 0.206 & 0.95 & 0.036 & 0.81 & 1.00 & 1.00 \\
         medium & DLIM-linear & 0.514 & 0.69 & 0.048 & 0.62 & 0.98 & -\\ 
         & DLM & 0.494 & 0.55 & 0.063 & 0.33 & -& -\\
         \hline
         
         & DLIM(20,20) & 0.025 & 0.92 & 0.010 & 0.91 & 1.00 & 1.00 \\
         high & DLIM-linear & 0.470 & 0.58 & 0.045 & 0.55 & 1.00 & -\\
         & DLM & 0.469 & 0.38 & 0.062 & 0.23 & -& -\\
         \hline

         \multicolumn{8}{c}{Scenario 4: Complex Modification} \\ \hline
         & DLIM(20,20) & 0.320 & 0.94 & 0.037 & 0.82 & 1.00 & 1.00 \\
         low & DLIM-linear & 0.481 & 0.79 & 0.040 & 0.74 & 0.52 & - \\ 
         & DLM & 0.442 & 0.71 & 0.049 & 0.52& -& - \\
         \hline

         & DLIM(20,20) & 0.163 & 0.95 & 0.028 & 0.81 & 1.00 & 1.00 \\ 
         medium & DLIM-linear & 0.395 & 0.69 & 0.036 & 0.63 & 0.98 & -\\ 
         & DLM & 0.394 & 0.54 & 0.047 & 0.35 & -& -\\
         \hline
         
         & DLIM(20,20) & 0.020 & 0.92 & 0.008 & 0.91 & 1.00 & 1.00 \\ 
         high & DLIM-linear & 0.359 & 0.59 & 0.034 & 0.55 & 1.00 & -\\ 
         & DLM & 0.374 & 0.42 & 0.046 & 0.24 & -& -\\  
         \hline
         \bottomrule
         
    \end{tabular} 
    
    \label{tab: main34}   
\end{table}

For simulation scenario 1, in which each individual's exposure-time-response function does not depend on its modifier value, all models performed similarly to each other. The standard DLM had cumulative RMSE lower than the DLIMs for all signal-to-noise settings, and slightly lower point-wise RMSE for the low signal-to-noise setting. We expected this because the simulated structure does not include modification to the true exposure-time-response functions. Our results demonstrate that when there is no modification, the DLM is slightly more efficient than both DLIMs. However, the gain in efficiency is relatively small.

For scenario 2, in which each individual's exposure-time-response function is linearly scaled by their modifying value, the DLIMs generally performed better than the standard DLM, which is mis-specified in this scenario. In the high signal-to-noise setting of scenario 2, the DLIM(20,20) and DLIM-linear had much lower cumulative RMSE (0.017 and 0.016, respectively) than the standard DLM (0.361). The DLIM(20,20) and DLIM-linear had much lower point-wise RMSE (0.003 and 0.002, respectively) than the standard DLM's point-wise RMSE (0.014). The DLIM(20,20) and DLIM-linear also had cumulative coverage (0.95) at the nominal level; whereas, the standard DLM's cumulative coverage (0.06) was well below the nominal level. The DLIM(20,20) and DLIM-linear also had point-wise coverage (0.98 and 0.94, respectively) near the nominal level; whereas, the standard DLM's point-wise coverage (0.55) was well below the nominal level. Our results demonstrate that when the signal-to-noise ratio is high, a model accounting for linear interaction in the estimated exposure-time-response function is more favorable than the standard DLM. In the medium and low signal-to-noise settings for scenario 2, the DLIMs out-performed the standard DLM in terms of cumulative effect RMSE and coverage. For point-wise metrics, all three models performed similarly in terms of RMSE, but the DLIM(20,20) and DLIM-linear had coverage near the nominal level (0.90 and 0.89, respectively) while the standard DLM had coverage lower than the nominal level (0.83). Our results show that if true modification to the exposure-time-response function is linear, then it is more appropriate to fit a DLIM. Fitting a standard DLM can result in low coverage.  

For the third and fourth simulation scenarios, in which each individual's exposure-time-response function also depends on its modifier value, a DLIM(20,20) was more accurate than the other models. For all signal-to-noise settings, the DLIM(20,20) had lower cumulative and point-wise RMSE than the other models. The DLIM(20,20) also had cumulative and point-wise coverage much closer to the nominal level than the other models. For example, cumulative coverage in the high signal-to-noise setting for scenario 3 was 0.92 for the DLIM(20,20); whereas, the cumulative coverage was 0.58 for the DLIM-linear and 0.38 for the standard DLM in the same setting. In the lower signal-to-noise ratio settings, point-wise coverage for the DLIM(20,20) model is closer to the nominal level than the other models, but is below the nominal level. For example in the low signal-to-noise setting for scenario 4, the point-wise coverage was 0.82; whereas, the point-wise coverage for the DLIM-linear was 0.74 and 0.52 for the standard DLM. We attribute the low point-wise coverage for these scenarios to attenuated estimates of the peak effect and over-smoothing of modification patterns in the low signal-to-noise settings. This persisted across various choices of basis constructions, number of basis functions, and penalty structures (see Supplemental Tables 1-6). As the signal-to-noise increases, our model can better estimate the peak effects and over-smoothing is reduced, resulting in smaller RMSE and coverage closer to the nominal level. Our results demonstrate that when modification to the exposure-time-response function is present, a DLIM(20,20) can accurately capture the complex modification patterns.

\subsubsection{Testing Modification}

Using the proposed bootstrap test, we were able to accurately identify scenarios with modification and without modification in the simulation study. Tables~\ref{tab: main12} and \ref{tab: main34} show the proportion of times that the null model was rejected at the $\alpha=0.05$ level using the test. For scenario 1 with no modification present, the tests comparing the DLIM(20,20) to the standard DLM and comparing the DLIM-linear to the standard DLM both had type I error rates between 0.04 and 0.06 across all signal-to-noise ratios. For scenario 2 with linear modification, the test for comparing the DLIM-linear to the standard DLM had power of 0.30 in the low signal setting, 0.80 in the medium signal setting, and 1.00 in the high signal setting. Additionally, the test for comparing the DLIM(20,20) to the DLIM-linear maintained a type I error rate near the nominal level, ranging from 0.04 to 0.06 across the signal-to-noise levels. For scenarios 3 and 4, the test accurately identified the DLIM(20,20) as the preferred model. The power for testing the DLIM(20,20) compared to both the DLIM-linear and standard DLM was 1.00 across all signal-to-noise settings.

\subsection{Hyperparameter Selection and additional simulations} \label{sec.tuning}

The DLIMs with PS(2,2) penalization generally performed better than DLIMs with alternative penalization methods discussed in Section~\ref{sec: penalty}. In particular, the DLIM(20,20) with PS(2,2) penalization performed better than the DLIM(20,20) with CR, PS(1,2), or PS(2,1) penalization. See Supplemental Tables 5 and 6 for a comparison to the PS(2,2) penalization method used in the simulation study. 

The penalized DLIMs generally performed better with a larger number of basis functions. Overall, our simulation study shows that using a large number of basis functions, e.g., 20, for both the exposure-time and modifier dimensions is best. Both the penalized DLM and penalized DLIM-linear did not benefit from an increased number of basis functions; therefore, we only include results for these models with 10 nominal degrees of freedom in the exposure-time dimension. All non-penalized models performed poorly and were sensitive to the choice of degrees of freedom. See Supplemental Tables 1-4 for a comparison between DLIMs fit with few and many basis functions and for non-penalized model results.

We considered an additional simulation with a normally distributed modifier. We present results in Supplemental Table 7, which were consistent with results presented in the main text.

We also demonstrate performance for a GLM version of the model. We present results for a simulation using count data in Supplemental Table 8, which were similar to results in Section~\ref{sec.parameter estimation results}
and suggest the use of our proposed model over the standard DLM.

\section{Analysis of Colorado Birth Cohort Data} \label{analysis}

\subsection{Data and Model Fitting}

We applied our penalized model to the Colorado birth cohort data set. Our analysis included BWGAZ as the response, weekly exposure to ambient $\text{PM}_{2.5}$ for the first 37 weeks of gestation as the exposure of interest, and all covariates described in Section \ref{sec.data}. In addition, we include categorical variables for maternal county of residence, and year and month of conception \citep{leung_bias_2023}. We used the Colorado EnviroScreen health and social factors score as the modifier. Health and social factors scores quantify neighborhood-level biological susceptibility and social vulnerabilities. Variables for race, ethnicity, and education are included into our model on both the individual level as covariates and on the neighborhood level as part of the health and social factors score as described in Section \ref{sec.data}. 

We fit a penalized DLM with nominal 20 degrees of freedom, a penalized DLIM-linear with nominal 20 degrees of freedom, and a penalized DLIM(20,20) and performed the model comparison test described in Section~\ref{sec: compare}. We performed all analyses using our \texttt{dlim} package in \texttt{R} version 4.3.0.

\subsection{Results}

\begin{figure}
    \centering
    \includegraphics[width=1\textwidth]{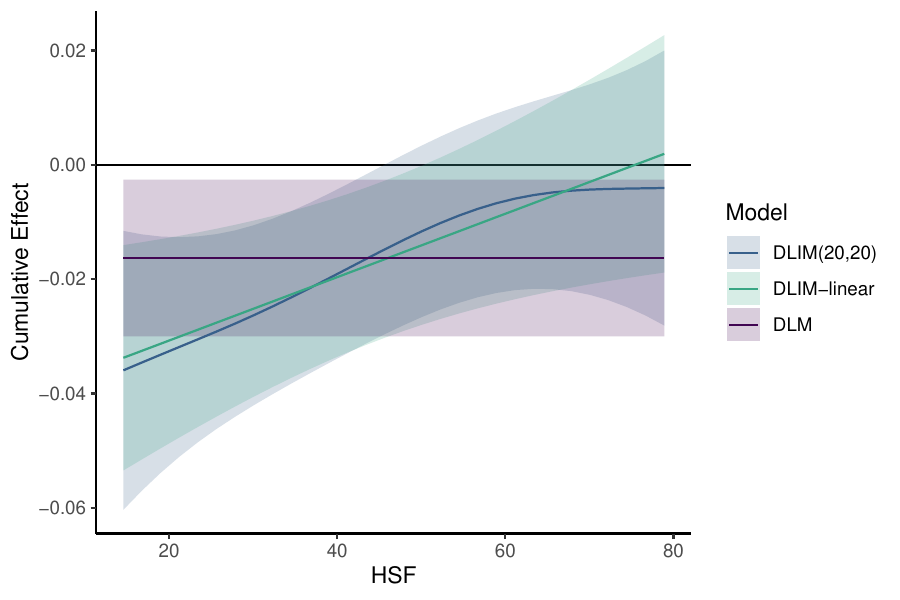}
    \caption{Estimates of cumulative effects of ambient $\text{PM}_{2.5}$ exposure during gestation on BWGAZ from a standard DLM, DLIM-linear, and DLIM(20,20) across the middle 98\% of Colorado EnviroScreen health and social factors (HSF) scores. We include confidence bands of 95\% around all estimates.}
    \label{fig:cumul_v2}
\end{figure}

Model comparisons with the bootstrap test supported  using the DLIM-linear. When comparing the DLIM-linear to the DLM, the empirical bootstrap p-value from the test was 0.019, indicating strong support for modification. The empirical bootstrap p-value was 0.077 when comparing the DLIM(20,20) to the DLM and was 0.402 when comparing the DLIM(20,20) to the DLIM-linear. This indicates that the data does not support the need for the more flexible DLIM(20,20) over the DLIM-linear. Furthermore, the results from the DLIM(20,20) show evidence of modification consistent with the DLIM-linear. Figure~\ref{fig:cumul_v2} shows the estimated cumulative effect by individual health and social factors score using each method. Cumulative effect estimates from the DLIM(20,20) are similar to those from the DLIM-linear, further supporting that the added flexibility of the DLIM(20,20) is not needed. Therefore, we chose the DLIM-linear for the analysis. See Supplemental Figure 1 for plots of point-wise effect estimates from the DLIM(20,20). 

Figure~\ref{fig:cumul_v2} shows the cumulative effect for each individual health and social factors score using each method. With a standard DLM, we estimated a negative cumulative effect of ambient $\text{PM}_{2.5}$ on BWGAZ for all values of the health and social factors score. With a DLIM-linear, we estimated a negative effect only for low health and social factors scores. Specifically, we identified an association between birth weight and increased maternal exposure to ambient $\text{PM}_{2.5}$ for health and social factors scores lower than 50. The cumulative effect for those health and social factors scores ranged from a decrease of 0.014 to 0.034 in BWGAZ for a 1 $\mu$g/$\text{m}^3$ difference in ambient $\text{PM}_{2.5}$. Our results demonstrate that the effect, cumulative across gestation, of ambient $\text{PM}_{2.5}$ on BWGAZ may be modified by neighborhood-level vulnerability of the population in an area.

\begin{figure}
    \centering
    \begin{subfigure}[b]{1\textwidth}
        \centering
        \includegraphics[width=1\textwidth]{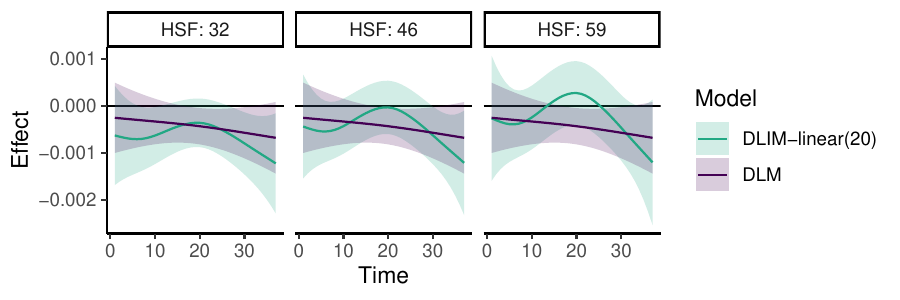}
        \label{subfig:pointwise_a_v2}
    \end{subfigure}
    \hfill
    \begin{subfigure}[b]{1\textwidth}
        \centering
        \includegraphics[width=1\textwidth]{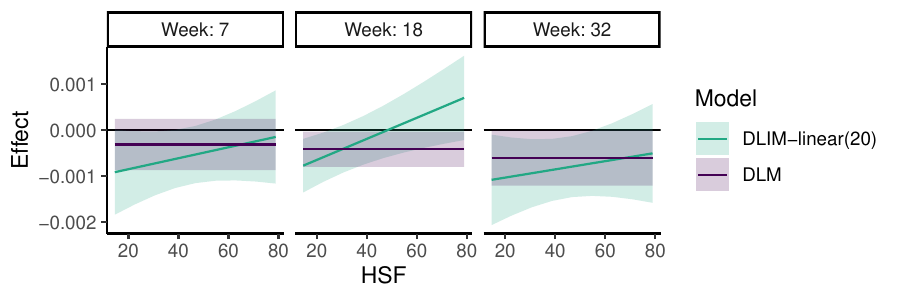}
        \label{subfig:pointwise_b_v2}
    \end{subfigure}
    \caption{Pointwise effect estimates of ambient $\text{PM}_{2.5}$ on BWGAZ from a standard DLM with 20 exposure-time degrees of freedom and a DLIM-linear with Colorado EnviroScreen health and social factors (HSF) score as a modifier. The top panel shows estimated exposure-time-response functions for each quartile of the HSF scores: scores 32, 46, and 59. The bottom panel shows estimated point-wise effects for a week around the middle of each trimester, excluding the first and last 1\% of HSF scores. We include confidence bands of 95\% around all estimates.}
    \label{fig:pointwise_v2}
\end{figure}

The top panel in Figure~\ref{fig:pointwise_v2} shows the estimated exposure-time-response functions for the 25th, 50th and 75th percentiles of the health and social factors scores: 32, 46, and 59, respectively. The estimated exposure-time-response functions are the same across health and social factors scores for the standard DLM but vary for the DLIM-linear. For a health and social factors score of 32, we identified a window of susceptibility during the the first and third trimesters. Specifically, we identified a small window from weeks 7 to 11 and another window from weeks 27 to 37. The peak effect was a decrease of 0.0007 in BWGAZ during the first window and a decrease of 0.0012 in BWGAZ during the second window. The windows we identified are similar in timing to those in some previous work; however, results on the specific timing of the windows are highly variable across studies \citep{bosetti_ambient_2010, stieb_ambient_2012, mork_heterogeneous_2023, niu_association_2022, johnson_critical_2022, gong_method_2022}. 

The bottom panel of Figure~\ref{fig:pointwise_v2} shows the estimated linear effect of exposure as a function of health and social factors scores at approximately the middle of each trimester: weeks 7, 18, and 32. See Supplemental Figures 2-4 for plots at all time points. In Figure~\ref{fig:pointwise_v2}, the effect for each week is constant across health and social factors scores for the standard DLM but varies for the DLIM-linear. We estimated a negative effect for some health and social factors scores with the DLIM-linear. During the first trimester, we estimated a negative effect for people who live in areas with health and social factors between 18 and 37. We estimated a negative effect during the middle of the second trimester for people who live in areas with health and social factors scores less than 25, and we estimated a negative effect during the middle of the third trimester for people who live in areas with health and social factors scores less than 55. Our findings are also reflected in the top panel. For example, we estimated a negative effect for people in an area with a health and social factors score of 46 during week 32, which is shown in both the second plot of the top panel and the third plot in the bottom panel. We did not estimate an effect of ambient $\text{PM}_{2.5}$ on BWGAZ with the standard DLM for any values of the health and social factors score at any time during gestation. Our results provide evidence that the effect of ambient $\text{PM}_{2.5}$ on BWGAZ at specific weeks during gestation may be associated with the susceptibility and vulnerability of the population in an individual's area of residence.

Overall, we found a negative association between ambient $\text{PM}_{2.5}$ exposure and BWGAZ consistent with a large body of literature \citep{bosetti_ambient_2010, jacobs_association_2017, lakshmanan_associations_2015, stieb_ambient_2012, sram_ambient_2005}. Further, we estimated this effect among individuals living in census tracts with low health and social factors scores and identified a small window of susceptibility for individuals with low health and social factors scores. The modification pattern is contrary to the expected results, as a low health and social factors score is intended to indicate decreased vulnerability to environmental exposures. Our analysis is the first observational study to use the Colorado EnviroScreen data. Hence, there is not a body of work for comparison. We propose a few explanations. First, similar inverse relationships have been found in some studies. For example, \cite{millett_healthy_2016} reviewed literature supporting immigrant children having a health advantage over native-born children of the same socio-economic status, and \cite{chu_immigrant_2022} found that foreign-born black and Latina women's children can have higher birth weight than native-born women's children in the United States. Second, area-level modification may be present, but our modifier may not be weighted correctly or may be misspecified. Colorado EnviroScreen uses geometric means to weight indicators into component scores, which our area-level modifier may not correctly capture. Finally, our conditioning on full-term births and gestational age \citep{neophytou_educational_2021}, along with live-birth bias \citep{raz_live-birth_2018} could explain this inverse relationship.

\section{Discussion} \label{sec. discussion}

In this paper, we proposed a distributed lag interaction model (DLIM). The proposed model extends the DLM framework to allow for individualized exposure-time-response functions based on a continuous modifying variable. The DLIM allows for estimation of personalized exposure-time-response functions by smoothing over both the exposure-time dimension and modifier dimension and for formal testing for interaction. Our proposed model is versatile, can be fit as GLMs and in mixed models, and is computationally efficient. Fitting and interpreting our DLIM is user-friendly through the development of our \texttt{dlim} package in \texttt{R}. 

Our proposed DLIM framework fills an existing gap in the literature. Many researchers aim to estimate the effect of air pollution on birth outcomes modified by a single continuous covariate, e.g., maternal stress and neighborhood vulnerability. Without the DLIM methodology, researchers must partition the continuous modifier into categorical factors and fit separate DLMs or BDLIMs for each subgroup \citep{bose_prenatal_2017, bose_prenatal_2018, lee_prenatal_2018, niu_association_2022}. While this approach provides an approximation of the effect over each subgroup of a continuous modifier, researchers cannot obtain a unique estimate for any value of the modifier without fitting a DLIM. Such approaches also include the potential for bias due to misspecification resulting from categorizing a continuous variable. 

Through simulation, we considered the possible methods of modification explored by \cite{wilson_bayesian_2017} and verified that a DLIM can accurately estimate the exposure-time-response function and cumulative effect whether or not there is modification. A standard DLM is more efficient when no modification is present, but only by a small amount. When modification is present, a standard DLM is not able to accurately capture the modification structure, which leads to bias in both the exposure-time-response function and cumulative effect estimates. While the gain in efficiency is small when fitting a standard DLM if there is no modification, the risk of fitting a standard DLM when there is modification is large.

We applied a DLIM to analyze the effect of ambient $\text{PM}_{2.5}$ on BWGAZ in Colorado modified by Colorado EnviroScreen's health and social factors score, a quantification of neighborhood-level vulnerability and susceptibility. We estimated a negative cumulative effect for lower health and social factors scores. This indicates that the susceptibility and vulnerability of pregnant people's areas of residence may modify the total effect of ambient $\text{PM}_{2.5}$ on BWGAZ. Our results provide evidence that pregnant people who live in areas that are less susceptible and vulnerable to social, economic, environmental, and climate adversities are at risk of having pregnancies with lower BWGAZ  when exposed to higher levels of ambient air pollution. 

Our DLIM framework is novel as it allows for smooth modification of an exposure-time-response function by a continuous factor. Existing methodology only allows for modification to the exposure-time-response function based on a categorical modifier. While our paper focuses on maternal exposure to air pollution and birth outcomes, the proposed methodology is broadly applicable in environmental health and other fields. For example, DLMs are often used in time series studies of air pollution on hospitalization or mortality rates \citep{schwartz_distributed_2000}, as well as in other fields \citep{schliep_distributed_2021, yang_effect_2022}.  An added level of novelty in this paper is using an index from Colorado EnviroScreen as a neighborhood-level modifying variable. Using this index as a modifier allows for modification based on 15 factors through a parsimonious index while fitting a single model that is computationally efficient and readily interpretable. 

\section*{Supplemental Material}

We reference additional tables and figures in Sections~\ref{sec.data},~\ref{sec. simulation},~and~\ref{analysis}. Code to replicate the simulations and the {\tt R} packages {\tt dlim} to implement the methods are available at 

\noindent https://github.com/ddemateis/dlim.

\section*{Acknowledgements}

National Institutes of Health grants ES029943 and ES030616 supported this work. 

These data were supplied by the Center for Health and Environmental Data Vital Statistics Program of the Colorado Department of Public Health and Environment, which specifically disclaims responsibility for any analyses, interpretations, or conclusions it has not provided.

\section*{Data availability statement}

The health data that support the findings in this paper are available from Colorado Department of Public Health and Environment (CDPHE) (https://cdphe.colorado.gov). Restrictions apply to the availability of these data, which were used under license in this paper. 

The exposure data that support the findings in this paper are available from the US EPA Air Data Repository (https://www.epa.gov/outdoor-air-quality-data) \citep{USAgency}  and the Community Multiscale Air Quality Modeling System processed with Fused Air Quality Surface Downscaler (https://www.epa.gov/hesc/rsig-related-downloadable-data-files) \citep{USAgency2, Berrocal2010AModels}. 

\bibliography{reference}

\begin{thebibliography}{}

\bibitem[Abatzoglou, 2013]{abatzoglou_development_2013}
Abatzoglou, J.~T. (2013).
\newblock Development of gridded surface meteorological data for ecological applications and modelling.
\newblock {\em International Journal of Climatology}, 33(1):121--131.

\bibitem[Arcaya et~al., 2016]{arcaya_research_2016}
Arcaya, M.~C., Tucker-Seeley, R.~D., Kim, R., Schnake-Mahl, A., So, M., and Subramanian, S. (2016).
\newblock Research on neighborhood effects on health in the {United} {States}: {A} systematic review of study characteristics.
\newblock {\em Social Science \& Medicine}, 168:16--29.

\bibitem[Armstrong, 2006]{armstrong_models_2006}
Armstrong, B. (2006).
\newblock Models for the {Relationship} {Between} {Ambient} {Temperature} and {Daily} {Mortality}.
\newblock {\em Epidemiology}, 17(6):624--631.

\bibitem[Becklake and Kauffmann, 1999]{becklake_gender_1999}
Becklake, M.~R. and Kauffmann, F. (1999).
\newblock Gender differences in airway behaviour over the human life span.
\newblock {\em Thorax}, 54(12):1119--1138.

\bibitem[Berrocal et~al., 2010]{Berrocal2010AModels}
Berrocal, V.~J., Gelfand, A.~E., and Holland, D.~M. (2010).
\newblock {A spatio-temporal downscaler for output from numerical models}.
\newblock {\em Journal of Agricultural, Biological, and Environmental Statistics}, 15(2):176--197.

\bibitem[Bose et~al., 2017]{bose_prenatal_2017}
Bose, S., Chiu, Y.-H.~M., Hsu, H.-H.~L., Di, Q., Rosa, M.~J., Lee, A., Kloog, I., Wilson, A., Schwartz, J., Wright, R.~O., Cohen, S., Coull, B.~A., and Wright, R.~J. (2017).
\newblock Prenatal {Nitrate} {Exposure} and {Childhood} {Asthma}. {Influence} of {Maternal} {Prenatal} {Stress} and {Fetal} {Sex}.
\newblock {\em American Journal of Respiratory and Critical Care Medicine}, 196(11):1396--1403.

\bibitem[Bose et~al., 2018]{bose_prenatal_2018}
Bose, S., Rosa, M.~J., Mathilda~Chiu, Y.-H., Leon~Hsu, H.-H., Di, Q., Lee, A., Kloog, I., Wilson, A., Schwartz, J., Wright, R.~O., Morgan, W.~J., Coull, B.~A., and Wright, R.~J. (2018).
\newblock Prenatal nitrate air pollution exposure and reduced child lung function: {Timing} and fetal sex effects.
\newblock {\em Environmental Research}, 167:591--597.

\bibitem[Bosetti et~al., 2010]{bosetti_ambient_2010}
Bosetti, C., Nieuwenhuijsen, M.~J., Gallus, S., Cipriani, S., La~Vecchia, C., and Parazzini, F. (2010).
\newblock Ambient particulate matter and preterm birth or birth weight: a review of the literature.
\newblock {\em Archives of Toxicology}, 84(6):447--460.

\bibitem[Chang et~al., 2015]{chang_assessment_2015}
Chang, H.~H., Warren, J.~L., Darrow, L.~A., Reich, B.~J., and Waller, L.~A. (2015).
\newblock Assessment of critical exposure and outcome windows in time-to-event analysis with application to air pollution and preterm birth study.
\newblock {\em Biostatistics}, 16(3):509--521.

\bibitem[Chiu et~al., 2016]{chiu_prenatal_2016}
Chiu, Y.-H.~M., Hsu, H.-H.~L., Coull, B.~A., Bellinger, D.~C., Kloog, I., Schwartz, J., Wright, R.~O., and Wright, R.~J. (2016).
\newblock Prenatal particulate air pollution and neurodevelopment in urban children: {Examining} sensitive windows and sex-specific associations.
\newblock {\em Environment International}, 87:56--65.

\bibitem[Chiu et~al., 2017]{chiu_prenatal_2017}
Chiu, Y.-H.~M., Hsu, H.-H.~L., Wilson, A., Coull, B.~A., Pendo, M.~P., Baccarelli, A., Kloog, I., Schwartz, J., Wright, R.~O., Taveras, E.~M., and Wright, R.~J. (2017).
\newblock Prenatal particulate air pollution exposure and body composition in urban preschool children: {Examining} sensitive windows and sex-specific associations.
\newblock {\em Environmental Research}, 158:798--805.

\bibitem[Chu et~al., 2022]{chu_immigrant_2022}
Chu, M.~T., Ettinger~de Cuba, S., Fabian, M.~P., Lane, K.~J., James-Todd, T., Williams, D.~R., Coull, B.~A., Carnes, F., Massaro, M., Levy, J.~I., Laden, F., Sandel, M., Adamkiewicz, G., and Zanobetti, A. (2022).
\newblock The immigrant birthweight paradox in an urban cohort: {Role} of immigrant enclaves and ambient air pollution.
\newblock {\em Journal of Exposure Science \& Environmental Epidemiology}, 32(4):571--582.

\bibitem[{Colorado Department of Public Health and Environment (CDPHE)}, 2022]{colorado_department_of_public_health_and_environment_cdphe_colorado_2022}
{Colorado Department of Public Health and Environment (CDPHE)} (2022).
\newblock Colorado {EnviroScreen} tool.
\newblock Technical report.
\newblock Publication Title: Colorado EnviroScreen Tool.

\bibitem[{Colorado EnviroScreen Tool Team}, 2022]{colorado_EnviroScreen_tool_team_colorado_2022}
{Colorado EnviroScreen Tool Team} (2022).
\newblock Colorado {EnviroScreen} tool technical document.
\newblock Technical report, Colorado EnviroScreen tool team., Denver, CO, USA.

\bibitem[Fenton, 2003]{fenton_new_2003}
Fenton, T.~R. (2003).
\newblock A new growth chart for preterm babies: {Babson} and {Benda}'s chart updated with recent data and a new format.
\newblock {\em BMC Pediatrics}, 3(1):13.

\bibitem[Gasparrini et~al., 2010]{gasparrini_distributed_2010}
Gasparrini, A., Armstrong, B., and Kenward, M.~G. (2010).
\newblock Distributed lag non-linear models.
\newblock {\em Statistics in Medicine}, 29(21):2224--2234.

\bibitem[Gasparrini et~al., 2017]{gasparrini_penalized_2017}
Gasparrini, A., Scheipl, F., Armstrong, B., and Kenward, M.~G. (2017).
\newblock A penalized framework for distributed lag non-linear models: {Penalized} {DLNMs}.
\newblock {\em Biometrics}, 73(3):938--948.

\bibitem[Gong and Zhan, 2022]{gong_method_2022}
Gong, X. and Zhan, F.~B. (2022).
\newblock A method for identifying critical time windows of maternal air pollution exposures associated with low birth weight in offspring using massive geographic data.
\newblock {\em Environmental Science and Pollution Research}, 29(22):33345--33360.

\bibitem[Green and Silverman, 1994]{green_nonparametric_1994}
Green, P.~J. and Silverman, B.~W. (1994).
\newblock {\em Nonparametric regression and generalized linear models: a rougnhness penalty approch}.
\newblock Number~58 in Monographs on statistics and applied probability. Chapman \& Hall, London Glasgow New York [etc.].

\bibitem[Hsu et~al., 2015]{hsu_prenatal_2015}
Hsu, H.-H.~L., Mathilda~Chiu, Y.-H., Coull, B.~A., Kloog, I., Schwartz, J., Lee, A., Wright, R.~O., and Wright, R.~J. (2015).
\newblock Prenatal {Particulate} {Air} {Pollution} and {Asthma} {Onset} in {Urban} {Children}. {Identifying} {Sensitive} {Windows} and {Sex} {Differences}.
\newblock {\em American Journal of Respiratory and Critical Care Medicine}, 192(9):1052--1059.

\bibitem[Jacobs et~al., 2017]{jacobs_association_2017}
Jacobs, M., Zhang, G., Chen, S., Mullins, B., Bell, M., Jin, L., Guo, Y., Huxley, R., and Pereira, G. (2017).
\newblock The association between ambient air pollution and selected adverse pregnancy outcomes in {China}: {A} systematic review.
\newblock {\em Science of The Total Environment}, 579:1179--1192.

\bibitem[Johnson et~al., 2022]{johnson_critical_2022}
Johnson, M., Shin, H.~H., Roberts, E., Sun, L., Fisher, M., Hystad, P., Van~Donkelaar, A., Martin, R.~V., Fraser, W.~D., Lavigne, E., Clark, N., Beaulac, V., and Arbuckle, T.~E. (2022).
\newblock Critical {Time} {Windows} for {Air} {Pollution} {Exposure} and {Birth} {Weight} in a {Multicity} {Canadian} {Pregnancy} {Cohort}.
\newblock {\em Epidemiology}, 33(1):7--16.

\bibitem[Laird and Ware, 1982]{laird_random-effects_1982}
Laird, N.~M. and Ware, J.~H. (1982).
\newblock Random-{Effects} {Models} for {Longitudinal} {Data}.
\newblock {\em Biometrics}, 38(4):963.

\bibitem[Lakshmanan et~al., 2015]{lakshmanan_associations_2015}
Lakshmanan, A., Chiu, Y.-H.~M., Coull, B.~A., Just, A.~C., Maxwell, S.~L., Schwartz, J., Gryparis, A., Kloog, I., Wright, R.~J., and Wright, R.~O. (2015).
\newblock Associations between prenatal traffic-related air pollution exposure and birth weight: {Modification} by sex and maternal pre-pregnancy body mass index.
\newblock {\em Environmental Research}, 137:268--277.

\bibitem[Lee et~al., 2018]{lee_prenatal_2018}
Lee, A., Leon~Hsu, H.-H., Mathilda~Chiu, Y.-H., Bose, S., Rosa, M.~J., Kloog, I., Wilson, A., Schwartz, J., Cohen, S., Coull, B.~A., Wright, R.~O., and Wright, R.~J. (2018).
\newblock Prenatal fine particulate exposure and early childhood asthma: {Effect} of maternal stress and fetal sex.
\newblock {\em Journal of Allergy and Clinical Immunology}, 141(5):1880--1886.

\bibitem[Lee et~al., 2020]{lee_prenatal_2020}
Lee, A.~G., Cowell, W., Kannan, S., Ganguri, H.~B., Nentin, F., Wilson, A., Coull, B.~A., Wright, R.~O., Baccarelli, A., Bollati, V., and Wright, R.~J. (2020).
\newblock Prenatal particulate air pollution and newborn telomere length: {Effect} modification by maternal antioxidant intakes and infant sex.
\newblock {\em Environmental Research}, 187:109707.

\bibitem[Leung et~al., 2023]{leung_bias_2023}
Leung, M., Rowland, S.~T., Coull, B.~A., Modest, A.~M., Hacker, M.~R., Schwartz, J., Kioumourtzoglou, M.-A., Weisskopf, M.~G., and Wilson, A. (2023).
\newblock Bias {Amplification} and {Variance} {Inflation} in {Distributed} {Lag} {Models} {Using} {Low}-{Spatial}-{Resolution} {Data}.
\newblock {\em American Journal of Epidemiology}, 192(4):644--657.

\bibitem[Marra and Wood, 2012]{marra_coverage_2012}
Marra, G. and Wood, S.~N. (2012).
\newblock Coverage {Properties} of {Confidence} {Intervals} for {Generalized} {Additive} {Model} {Components}: {Coverage} properties of {GAM} intervals.
\newblock {\em Scandinavian Journal of Statistics}, 39(1):53--74.

\bibitem[Martenies et~al., 2022a]{martenies_using_2022}
Martenies, S.~E., Hoskovec, L., Wilson, A., Moore, B.~F., Starling, A.~P., Allshouse, W.~B., Adgate, J.~L., Dabelea, D., and Magzamen, S. (2022a).
\newblock Using non-parametric {Bayes} shrinkage to assess relationships between multiple environmental and social stressors and neonatal size and body composition in the {Healthy} {Start} cohort.
\newblock {\em Environmental Health}, 21(1):111.

\bibitem[Martenies et~al., 2022b]{martenies_associations_2022}
Martenies, S.~E., Zhang, M., Corrigan, A.~E., Kvit, A., Shields, T., Wheaton, W., Bastain, T.~M., Breton, C.~V., Dabelea, D., Habre, R., Magzamen, S., Padula, A.~M., Him, D.~A., Camargo, C.~A., Cowell, W., Croen, L.~A., Deoni, S., Everson, T.~M., Hartert, T.~V., Hipwell, A.~E., McEvoy, C.~T., Morello-Frosch, R., O'Connor, T.~G., Petriello, M., Sathyanarayana, S., Stanford, J.~B., Woodruff, T.~J., Wright, R.~J., and Kress, A.~M. (2022b).
\newblock Associations between combined exposure to environmental hazards and social stressors at the neighborhood level and individual perinatal outcomes in the {ECHO}-wide cohort.
\newblock {\em Health \& Place}, 76:102858.

\bibitem[Millett, 2016]{millett_healthy_2016}
Millett, L.~S. (2016).
\newblock The {Healthy} {Immigrant} {Paradox} and {Child} {Maltreatment}: {A} {Systematic} {Review}.
\newblock {\em Journal of Immigrant and Minority Health}, 18(5):1199--1215.

\bibitem[Mork et~al., 2023]{mork_heterogeneous_2023}
Mork, D., Kioumourtzoglou, M.-A., Weisskopf, M., Coull, B.~A., and Wilson, A. (2023).
\newblock Heterogeneous {Distributed} {Lag} {Models} to {Estimate} {Personalized} {Effects} of {Maternal} {Exposures} to {Air} {Pollution}.
\newblock {\em Journal of the American Statistical Association}, pages 1--13.

\bibitem[Neophytou et~al., 2021]{neophytou_educational_2021}
Neophytou, A.~M., Kioumourtzoglou, M.-A., Goin, D.~E., Darwin, K.~C., and Casey, J.~A. (2021).
\newblock Educational note: addressing special cases of bias that frequently occur in perinatal epidemiology.
\newblock {\em International Journal of Epidemiology}, 50(1):337--345.

\bibitem[Niu et~al., 2022]{niu_association_2022}
Niu, Z., Habre, R., Chavez, T.~A., Yang, T., Grubbs, B.~H., Eckel, S.~P., Berhane, K., Toledo-Corral, C.~M., Johnston, J., Dunton, G.~F., Lerner, D., Al-Marayati, L., Lurmann, F., Pavlovic, N., Farzan, S.~F., Bastain, T.~M., and Breton, C.~V. (2022).
\newblock Association {Between} {Ambient} {Air} {Pollution} and {Birth} {Weight} by {Maternal} {Individual}- and {Neighborhood}-{Level} {Stressors}.
\newblock {\em JAMA Network Open}, 5(10):e2238174.

\bibitem[Peng et~al., 2017]{peng_short-term_2017}
Peng, C., Sanchez-Guerra, M., Wilson, A., Mehta, A.~J., Zhong, J., Zanobetti, A., Brennan, K., Dereix, A.~E., Coull, B.~A., Vokonas, P., Schwartz, J., and Baccarelli, A.~A. (2017).
\newblock Short-term effects of air temperature and mitochondrial {DNA} lesions within an older population.
\newblock {\em Environment International}, 103:23--29.

\bibitem[Raz et~al., 2018]{raz_live-birth_2018}
Raz, R., Kioumourtzoglou, M.-A., and Weisskopf, M.~G. (2018).
\newblock Live-{Birth} {Bias} and {Observed} {Associations} {Between} {Air} {Pollution} and {Autism}.
\newblock {\em American Journal of Epidemiology}, 187(11):2292--2296.

\bibitem[Roca-Pardiñas et~al., 2005]{roca-pardinas_testing_2005}
Roca-Pardiñas, J., Cadarso-Suárez, C., and González-Manteiga, W. (2005).
\newblock Testing for interactions in generalized additive models: {Application} to {SO2} pollution data.
\newblock {\em Statistics and Computing}, 15(4):289--299.

\bibitem[Schliep et~al., 2021]{schliep_distributed_2021}
Schliep, E.~M., Schafer, T. L.~J., and Hawkey, M. (2021).
\newblock Distributed lag models to identify the cumulative effects of training and recovery in athletes using multivariate ordinal wellness data.
\newblock {\em Journal of Quantitative Analysis in Sports}, 17(3):241--254.

\bibitem[Schnake-Mahl et~al., 2020]{schnake-mahl_gentrification_2020}
Schnake-Mahl, A.~S., Jahn, J.~L., Subramanian, S., Waters, M.~C., and Arcaya, M. (2020).
\newblock Gentrification, {Neighborhood} {Change}, and {Population} {Health}: a {Systematic} {Review}.
\newblock {\em Journal of Urban Health}, 97(1):1--25.

\bibitem[Schwartz, 2000]{schwartz_distributed_2000}
Schwartz, J. (2000).
\newblock The {Distributed} {Lag} between {Air} {Pollution} and {Daily} {Deaths}:.
\newblock {\em Epidemiology}, 11(3):320--326.

\bibitem[Stieb et~al., 2012]{stieb_ambient_2012}
Stieb, D.~M., Chen, L., Eshoul, M., and Judek, S. (2012).
\newblock Ambient air pollution, birth weight and preterm birth: {A} systematic review and meta-analysis.
\newblock {\em Environmental Research}, 117:100--111.

\bibitem[{US Environmental Protection Agency}, 2020a]{USAgency}
{US Environmental Protection Agency} (2020a).
\newblock {Air Quality System Data Mart [internet database]}.
\newblock {\em Available via https://www.epa.gov/airdata. Accessed June 01, 2020}.

\bibitem[{US Environmental Protection Agency}, 2020b]{USAgency2}
{US Environmental Protection Agency} (2020b).
\newblock {Fused Air Quality Surface Using Downscaling (FAQSD) Files}.
\newblock {\em Available via https://www.epa.gov/hesc/rsig-related-downloadable-data-files. Accessed June 01, 2020}.

\bibitem[Warren et~al., 2012]{warren_spatial-temporal_2012}
Warren, J., Fuentes, M., Herring, A., and Langlois, P. (2012).
\newblock Spatial-{Temporal} {Modeling} of the {Association} between {Air} {Pollution} {Exposure} and {Preterm} {Birth}: {Identifying} {Critical} {Windows} of {Exposure}.
\newblock {\em Biometrics}, 68(4):1157--1167.

\bibitem[Warren et~al., 2013]{warren_air_2013}
Warren, J.~L., Fuentes, M., Herring, A.~H., and Langlois, P.~H. (2013).
\newblock Air {Pollution} {Metric} {Analysis} {While} {Determining} {Susceptible} {Periods} of {Pregnancy} for {Low} {Birth} {Weight}.
\newblock {\em ISRN Obstetrics and Gynecology}, 2013:1--9.

\bibitem[Warren et~al., 2020]{warren_spatially_2020}
Warren, J.~L., Luben, T.~J., and Chang, H.~H. (2020).
\newblock A spatially varying distributed lag model with application to an air pollution and term low birth weight study.
\newblock {\em Journal of the Royal Statistical Society: Series C (Applied Statistics)}, 69(3):681--696.

\bibitem[Wilson et~al., 2017a]{wilson_bayesian_2017}
Wilson, A., Chiu, Y.-H.~M., Hsu, H.-H.~L., Wright, R.~O., Wright, R.~J., and Coull, B.~A. (2017a).
\newblock Bayesian distributed lag interaction models to identify perinatal windows of vulnerability in children’s health.
\newblock {\em Biostatistics}, 18(3):537--552.

\bibitem[Wilson et~al., 2017b]{wilson_potential_2017}
Wilson, A., Chiu, Y.-H.~M., Hsu, H.-H.~L., Wright, R.~O., Wright, R.~J., and Coull, B.~A. (2017b).
\newblock Potential for {Bias} {When} {Estimating} {Critical} {Windows} for {Air} {Pollution} in {Children}’s {Health}.
\newblock {\em American Journal of Epidemiology}, 186(11):1281--1289.

\bibitem[Wood, 2017]{wood_generalized_2017}
Wood, S.~N. (2017).
\newblock {\em Generalized additive models: an introduction with {R}}.
\newblock Chapman \& {Hall}/{CRC} texts in statistical science. CRC Press/Taylor \& Francis Group, Boca Raton, second edition edition.

\bibitem[Yang et~al., 2022]{yang_effect_2022}
Yang, D., Chen, L., Yang, Y., Shi, J., Huang, Z., Li, M., Yang, Y., and Ji, X. (2022).
\newblock Effect of {PM2}.5 exposure on {Vitamin} {D} status among pregnant women: {A} distributed lag analysis.
\newblock {\em Ecotoxicology and Environmental Safety}, 239:113642.

\bibitem[Yip, 1987]{yip_altitude_1987}
Yip, R. (1987).
\newblock Altitude and birth weight.
\newblock {\em The Journal of Pediatrics}, 111(6):869--876.

\bibitem[Zanobetti, 2000]{zanobetti_generalized_2000}
Zanobetti, A. (2000).
\newblock Generalized additive distributed lag models: quantifying mortality displacement.
\newblock {\em Biostatistics}, 1(3):279--292.

\bibitem[Šrám et~al., 2005]{sram_ambient_2005}
Šrám, R.~J., Binková, B., Dejmek, J., and Bobak, M. (2005).
\newblock Ambient {Air} {Pollution} and {Pregnancy} {Outcomes}: {A} {Review} of the {Literature}.
\newblock {\em Environmental Health Perspectives}, 113(4):375--382.

\end{thebibliography}

\end{document}